\newcommand{\CLASSINPUTtoptextmargin}{19mm}%
\newcommand{\CLASSINPUTbottomtextmargin}{43mm}%
\newcommand{\CLASSINPUTinnersidemargin}{12.9mm}%
\newcommand{\CLASSINPUToutersidemargin}{12.9mm}%
\documentclass[journal,10pt,a4paper]{IEEEtran}% requires IEEEtran V1.8+
\usepackage{hyperref}
\usepackage{amsmath}% for double integral symbol in this template
\usepackage{times}% use times font for the paper instead of default Computer Modern fonts
\usepackage{graphicx}% for figures
\usepackage{multirow}% to allow multiple-row elements in tabular environment
\usepackage[none]{hyphenat}% turn off hyphenation to make text extraction and indexing easier
\usepackage{float}% better control of floating figures and tables
\usepackage{subfig}% for subfigures within figures
\usepackage{longtable}
\usepackage{tabularx,booktabs}
\usepackage{multirow}
\usepackage{makecell}
\usepackage{eqparbox}
\usepackage{mdwmath}
\usepackage{mdwtab}

\makeatletter
\def\ps@IEEEtitlepagestyle{%
  \def\@oddfoot{\mycopyrightnotice}%
}
\def\mycopyrightnotice{%
  \begin{minipage}{\textwidth}
  \centering \scriptsize
  \copyright 2023 IEEE. Personal use of this material is permitted. Permission from IEEE must be obtained for all other uses, in any current or future media, including reprinting/republishing this material for advertising or promotional purposes, creating new collective works, for resale or redistribution to servers or lists, or reuse of any copyrighted component of this work in other works.
  \end{minipage}
}
\makeatother
\begin{document}
%%%%%%%%%%%%%%%%%%%%%%%%%%%%%%%%%%%%%%%%%%%%%%%%%%%%%%%%%%%%%%%%%%%%%%%%%%%%%
% We use \raggedbottom to avoid latex adding vertical space around headings.
% This gives a better idea to the author about how much white space remains
% as the page limit is approached.
\raggedbottom
%
%%%%%%%%%%%%%%%%%%%%%%%%%%%%%%%%%%%%%%%%%%%%%%%%%%%%%%%%%%%%%%%%%%%%%%%%%%%%%
% PAPER TITLE AND AUTHOR BLOCK
%
% The paper title can use linebreaks \\ within to get better formatting if desired.
%
\title{RCS-based Quasi-Deterministic Ray Tracing for Statistical Channel Modeling}
%
% Next we define the author names and affiliations.
% Author names are listed using \IEEEauthorblockN{} with comma separators between names.
% Affiliations are listed using \IEEEauthorblock{} with \\ separators between affiliations.
% Symbols marking author-affiliation relations are output using \EuMWauthorrefmark{}.
% At the end of the affiliation list is the list of author emails.
% See below for examples of each of these.
%

\author{
Javad Ebrahimizadeh, Evgenii Vinogradov, Guy A.E. Vandenbosch \thanks{J. Ebrahimizadeh and G. Vandenbosch are with WaveCoRE of the Department of Electrical Engineering (ESAT), KU Leuven, Leuven, Belgium. E-mail: \{Javad.Ebrahimizade,Guy.Vandenbosch\}@kuleuven.be}
\thanks{E. Vinogradov is with ESAT, KU Leuven, Leuven, Belgium, also with Autonomous Robotics Research Center, Technology Innovation Institute (TII), Abu Dhabi, UAE. E-mail: evgenii.vinogradov@tii.ae.}
}

\maketitle

\begin{abstract}
This paper presents a quasi-deterministic ray tracing (QD-RT) method for analyzing the propagation of electromagnetic waves in street canyons. The method uses a statistical bistatic distribution to model the Radar Cross Section (RCS) of various irregular objects such as cars and pedestrians, instead of relying on exact values as in a deterministic propagation model. The performance of the QD-RT method is evaluated by comparing its generated path loss distributions to those of the deterministic ray tracing (D-RT) model using the Two-sample Cramer-von Mises test. The results indicate that the QD-RT method generates the same path loss distributions as the D-RT model while offering lower complexity. This study suggests that the QD-RT method has the potential to be used for analyzing complicated scenarios such as street canyon scenarios in mmWave wireless communication systems.
\end{abstract}

\begin{IEEEkeywords}
quasi-deterministic, ray tracing, Radar Cross Section, statistical distribution, EM  propagation, Cramer-von Mises test.
\end{IEEEkeywords}
%
%%%%%%%%%%%%%%%%%%%%%%%%%%%%%%%%%%%%%%%%%%%%%%%%%%%%%%%%%%%%%%%%%%%%%%%%%%%%%
% THE REST OF THE PAPER follows.
%

\section{Introduction}
Wireless communication has been rapidly evolving with the advent of new technologies and the increasing demand for high-speed data transmission. Millimeter-Wave (mmWave) wireless communication is considered a promising technology for the next generation of wireless communication due to its ability to provide multi-Gbps average data rates with low latency \cite{ghasempour2017ieee}. This high data rate is particularly necessary for dense urban areas such as the street canyon scenario, where a large number of users demand high-speed data transmission. In this scenario, radio frequencies at mmWave bands are used to transmit data, which requires an understanding of the propagation characteristics of mmWave signals in street canyons. Recently, Facebook introduced an affordable solution for deploying high-speed data access in street canyons using mmWave Terragraph radios operating at 60 GHz for rooftop-to-rooftop or light-pole-to-light-pole links \cite{du202260,dupleich2019multi}.

Since there is no closed-form scattering model available for bistatic  Radar Cross Section (RCS) of irregular objects such as pedestrians and cars, numerical methods such as the Method of Moments (MoM), Geometrical Optics (GO), Physical Optics (PO), or their combinations, are typically used to calculate the bistatic RCS  of these objects. However, this increases the computational complexity of the analysis, which can be especially challenging in the case of the street canyon scenario, where a large number of irregular objects need to be considered.

While the use of the bistatic RCS  model of a sphere in the METIS channel model is simple, it may not accurately represent the scattering from irregular objects in all directions. This is because a large sphere, relative to the wavelength, exhibits a constant RCS. To address this limitation, Lahuerta-Lavieja et al. developed a fast mmWave scattering model based on the 3D Fresnel model for rectangular surfaces. However, while these models are useful for certain types of objects, they may not accurately model more complex or irregular objects \cite{nurmela2015deliverable,lahuerta2022computationally}. Therefore, further research is needed to develop more accurate bistatic RCS  models that can be incorporated into channel models for a more comprehensive analysis of wireless communication systems in real-world scenarios. 

Myint et al. demonstrated the feasibility of modeling the bistatic RCS of intricate objects using a closed-form statistical distribution function. They found that the bistatic RCS of cars conforms to a logistic distribution and applied this model to various vehicle types, including passenger cars, vans, and trucks, at sub-6 GHz frequency. However, they did not validate this Probability Density Function (PDF) model in a practical channel environment \cite{myint2019statistical}.

The present  paper introduces a low-complexity quasi-deterministic ray tracing method that takes advantage of the statistical distribution of bistatic RCS of irregular objects for calculating scattering instead of its exact values, as done in deterministic ray tracing. The method uses the Physical Optics (PO) method to calculate the bistatic RCS of irregular objects at mmWave and assigns a suitable Probability Density Function (PDF) to them. This approach significantly reduces the complexity of the ray tracing method. The QD-RT method is verified numerically by calculating the path loss due to irregular objects in a realistic street canyon scenario. 

The main  contributions of the paper are:

\begin{itemize}
    \item  Development of  a quasi-deterministic ray tracing technique based on dedicated PDFs  of bistatic RCSs of  objects.
    \item  Deriving   the probability density function of the area coverage for a specific street canyon scenario in spherical coordinates. 
\end{itemize}
The rest of the paper is organized as follows. Section \ref{SecTheory} describes  the quasi-deterministic ray tracing method. Section \ref{SecSimResult}   validates the quasi-deterministic propagation technique.  Finally, the paper is concluded in Section \ref{SecConclusion}.
%%%%%%%%%%%%%%%%%%%%%%%%%%%%%%%%%%%%%%%%%%%%%%%%%%%%%%%%%%%%%%%%%%%%%%%%%%%%%

\section{quasi-deterministic ray tracing method}
\label{SecTheory}
In this section, we provide a comprehensive overview of the street canyon topology and the theory of deterministic electromagnetic (EM)  propagation in the scenario. Additionally, we outline the quasi-deterministic and statistical channel models used in the study and their corresponding parameterization.
%%%%%%%%%%%%%%%%%%%%%%%%%%%%%%%%%%%%%%%%%%%%%%%%%%%%%%%%%%%%%%%%%%%%%%%%%%%%%
\begin{figure}
\centering
       \subfloat[]{
          \includegraphics[width=6cm, height=4.5cm]{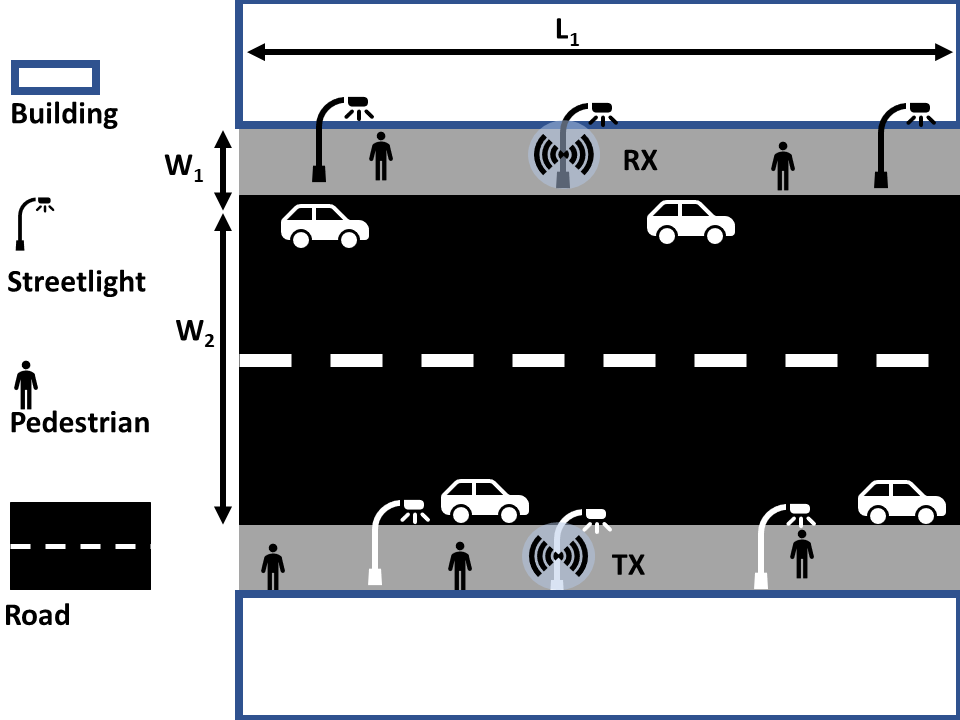}
          \label{u80}
       } 
       \vfill
       \subfloat[]{
          \includegraphics[width=6cm, height=4.5cm]{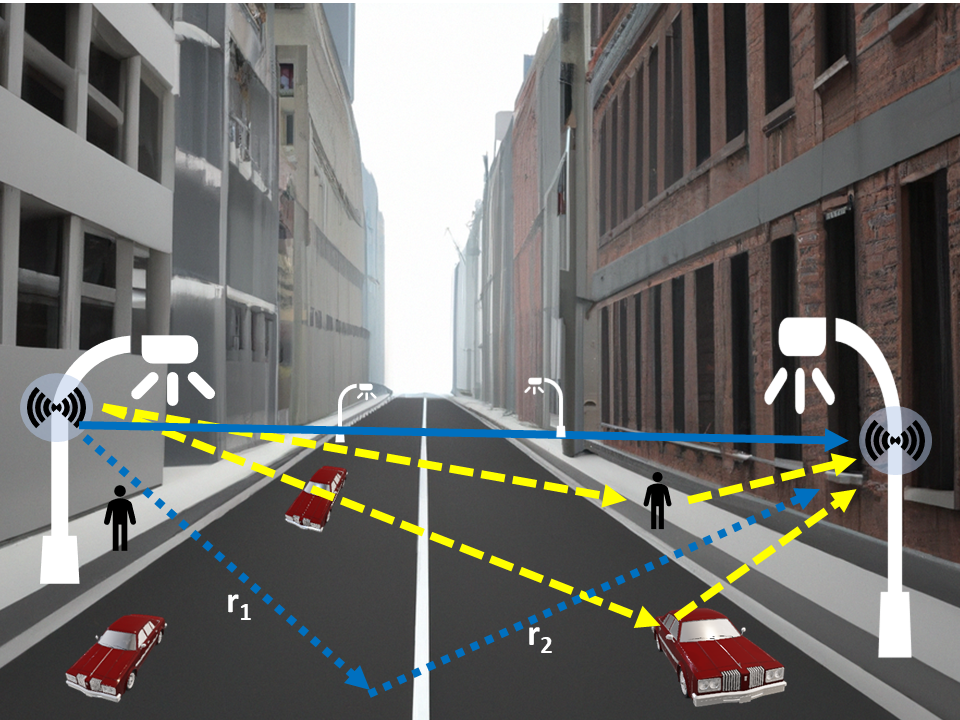}
          \label{u10}%
       }
\caption{Street canyon scenario topology which consists of Line of Sight, reflected scattered  paths. The solid line is LoS, dotted lines are reflected, and the dash-dotted line shows
scattered rays: (a) Top view (b) Perspective view.}
\label{Scenariotopology}
\end{figure}

\begin{table}[]
    \centering
     \caption{Overview of Simulation Scenario.}
     \label{SimulationSetupsummary}
    \begin{tabular}{|c|c|c|}
\hline
\thead{ \bfseries Object} & \bfseries Parameters &\bfseries{ Simulation values }   \\ \hline
 Frequency  & \makecell{$f_0$} &  60 GHz  \\ \hline
\makecell{TX and RX\\ antenna \\type} & Gain  & 0~[dBi] \\ \hline
Polarization& \makecell{Co-polarization} &  V-V  \\ \hline
\makecell{TX and RX \\antenna\\ locations} & \makecell{  $(x_{TX},y_{TX},z_{TX}) $\\$(x_{RX},y_{RX},z_{RX}) $ }  & \makecell{  $(0,2,3.5)~[m]$\\$(0,15,1.5)~[m]$ }  \\ \hline
\makecell{Pedestrian\\ model \\dimension}&  \makecell{   Length : $L_p $\\ Width : $W_p $\\ Height : $H_p $ }   &  \makecell{  0.4~[m] \\ 0.4~[m]\\ 1.8~[m]}   \\ \hline
\makecell{Parked car\\ model \\dimension}&   \makecell{   Length : $L_c$\\ Width : $W_c $\\ Height: $H_c $ }   &  \makecell{   4.55~[m]\\ 1.77~[m]\\ 1.24~[m]}   \\ \hline
Lamppost   &  \makecell{   Radius : $R_1 $\\ Length : $L_1 $\\ Separation distance : $d_1 $ \\ Total number: $N_1 $ \\ locations : $y_l $}   &  \makecell{   10~[cm] \\ 3~[m] \\  32~[m]  \\  10 \\ 2~[m] and 14~[m]}    \\ \hline
Sidewalk &  \makecell{  Length : $L_1 $\\ Width : $W_2 $}  &  \makecell{   150~[m] \\   2~[m]}\\ \hline
Street &  \makecell{  Length : $L_1 $\\ Width : $W_1 $}  &  \makecell{   150~[m] \\   12~[m] }\\ \hline
Wall &  \makecell{  Length : $L_w $\\ Thickness : $D_w $ \\ Height: $H_w$ \\ Relative permittivity  $\epsilon_{r,w}$ \\ locations: $Y_w$ }   &   \makecell{  150~[m] \\ 10~[cm]\\ Infinite \\ 3.26  \\ 0 and 16[m] }   \\ \hline
Ground &  \makecell{   Type   \\ Relative permittivity  $ \epsilon_{r,g}$}   &   \makecell{   Dry ground \\   6  }   \\ \hline
    \end{tabular}
    % \caption{Caption}
    % \label{tab:my_label}
\end{table}

\subsection{\uppercase{s}treet canyon scenario topology}
The topology of the street canyon scenario is shown in Figure \ref{Scenariotopology}  with two tall buildings on either side of the street. The street has a length of $W_2$ and a width of $L_1$, and there is a sidewalk on both sides of the street with a width of $W_1$. In this scenario, there are scattering objects such as lampposts, parked cars, and pedestrians placed on the street. The walls of the buildings have a thickness of $D_w$ and are made of bricks with a relative permittivity of $\epsilon_{r,w}$ at operational frequency $f_0$. The transmitter and receiver antennas are omnidirectional antennas with vertical polarization, and they are located at positions $(X_{tx}, Y_{tx}, Z_{tx})$ and $(X_{rx}, Y_{rx}, Z_{rx})$, respectively. The lampposts have a radius of $R_l$ and a length of $L_l$, and they are equidistantly positioned on both sides of the street with a separation distance of $d_l$. The scenario dimensions and parameter values are provided in Table \ref{SimulationSetupsummary}.
\subsection{\uppercase{d}eterministic propagation}
The propagation of Em wave in street canyon scenario includes the Line of Sight (LOS),  reflection, and scattering paths without considering shadowing, and diffraction. 
The LOS, reflection (from walls and ground), and scattering components can be modeled as:
\subsubsection{	LOS propagation}
\begin{equation}
\begin{aligned}
\label{LoS}
H_0(\omega)=a_0e^{j\omega \tau_0},
\end{aligned}
\end{equation}
where $\mid a_0\mid ^{2}=(\frac{\lambda}{4\pi r_0})^{2}$ is the LOS propagation loss,  the corresponding path loss in dB is  $PL=-20~log_{10}(|a_0|)$,  and $\tau_0=\frac{r_0}{c_0}$ is the propagation time. 

\subsubsection{reflections from ground and walls}
\begin{equation}
\begin{aligned}
\label{Ref}
H^r(\omega)=a^re^{j\omega \tau_r},
\end{aligned}
\end{equation}
where $\mid a^r\mid ^{2}=(\frac{R^{TE/TM} \lambda}{4\pi (r_1+r_2)})^{2}$, the corresponding   path loss  in dB due to reflection is  $PL=-20~log_{10}(|a^r|)$,   and $\tau^r=\frac{(r_1+r_2)}{c_0}$ is the propagation time;  $r_1$ is the  distance between TX to specular point and $r_2$ is the distance between the specular point to RX. The reflection coefficient   $R^{TE/TM}$ for both TE and TM polarization for dielectric slab (wall) and half-space (Ground) media  \cite{chew1999waves}.
\subsubsection{ Scattering from  objects}
\begin{equation}
\begin{aligned}
\label{Scat}
H^s(\omega)=a^se^{j\omega \tau_s},
\end{aligned}
\end{equation}
where   $\mid a^s\mid ^{2}= \frac{1}{4\pi r_1^2} \times  {\sigma_{rcs}}\times \frac{1}{4\pi r_2^2} \times \frac{\lambda^2}{4\pi} $,   the corresponding path loss in dB due to scattering  is  $PL=-20~log_{10}(|a^s|)$,  and the propagation time is $\tau^s=\frac{(r_1+r_2)}{c_0}$; $r_1$  and $r_2$  are the  distance between the scatterer and RX and TX, respectively and $\sigma_{rcs}$ is the bistatic RCS of the scatterer.  In this paper, the bistatic RCS values of complex objects (such as cars or pedestrians) are computed by the Physical Optics Gordon method, and regular shape objects (e.g., lampposts) are computed with the closed-form model of the RCS of a conducting cylinder \cite{balanis2012advanced}.  
\subsection{\uppercase{q}uasi-deterministic propagation}
In the quasi-deterministic ray tracing (QD-RT) method, the  PDF  of a bistatic RCS in   (\ref{Scat}) is used instead of the exact value of this bistatic RCS resulting in computational complexity decreases drastically. The QD-RT method is a low-complexity technique for statistically analysis and modeling of the channel with Monte-Carlo simulations should be done. A Monte Carlo simulation has variables that should be randomly varied during each iteration. For example for statistical analysis of the path loss due to an irregular object using  (\ref{Scat}), the distance between the object and the TX and RX antenna denoted by $r_1$ and $r_2$ are considered as the Monte-Carlo variable denoted  as the independent random variable  $X_1$ and   $X_2$.  Therefore, based on  (\ref{Scat}), the path loss is a random variable  denoted as
\begin{equation}
\begin{aligned}
\label{pathlossScat}
&PL(X_1,X_2) \sim  A_0-\\& 40\times log_{10}(X_1+X_2)-10\times log_{10}(\sigma_{rcs})
\end{aligned}
\end{equation}
where $A_0=-10log_{10}((4\pi)^3 \times \lambda^2)$ is a constant value. According to  (\ref{pathlossScat}), using the PDF of the bistatic RCS of objects can generate the same distribution for the path loss as using the exact values of bistatic RCS.

To model the bistatic Radar Cross Section (RCS) of an irregular object using a Probability Density Function (PDF), a dataset of bistatic RCS for all incident and scattered angles must be generated. It is important to note that the combination of the bistatic RCS at different angles to generate the dataset of bistatic RCS is not equal and depends on the specific scenario being tested. In the case of a street canyon scenario, the angular dependency of the bistatic RCS in creating the dataset follows a specific equation:
\begin{equation}
\begin{split}
\label{EqCoverageProb}
&f_{\Theta, \Phi}(\theta,\phi)=   \\ &
\begin{cases}
\frac{(\Delta z)^2sin(\theta)}{2L_1W_2~cos^3(\theta)} & \text{, if } 
\begin{cases}
    \frac{a}{\Delta z \times sin(\phi)}<\theta<\frac{b}{\Delta z \times sin(\phi)}\\ 
    \phi_0 < \phi < \pi - \phi_0,               
\end{cases}\\
0& , \mathrm{~otherwise}
\end{cases}
\end{split}
\end{equation}
where ($\theta$ and $\phi$) are elevation and azimuth angles in spherical coordinates. $\Delta z$ is the differential height between the object and TX (RX).    Here, the elevation angle is limited by the lines  $Y_w = a$ and  $Y_w = b$ and the azimuth angle  is bounded by  $\phi_0 = \frac{2a}{L_1}$, see Fig~.\ref{u10}.
\begin{table}[h]
\centering
\caption{Statistical parameters of logistic  distributions of bistatic RCSs.}
\label{bistaticRCSPDF}
\begin{tabular}{ |p{1.5cm}|p{1.5cm}|p{1.cm}|p{1.5cm}|   }
 \hline
 Object &  PDF    & Mean ($\mu $) [dBsm]    & Scale parameter  
 ($\sigma$) [dBsm] \\
 \hline
Pedestrian  & Logistic & 6.17 & 3.9     \\ \hline
Parked car  & Logistic & 11 & 4.3     \\ \hline
\end{tabular}
\end{table}
\begin{figure}[h]
  \begin{center}
  \includegraphics[width=8.3cm, height=5.5cm]{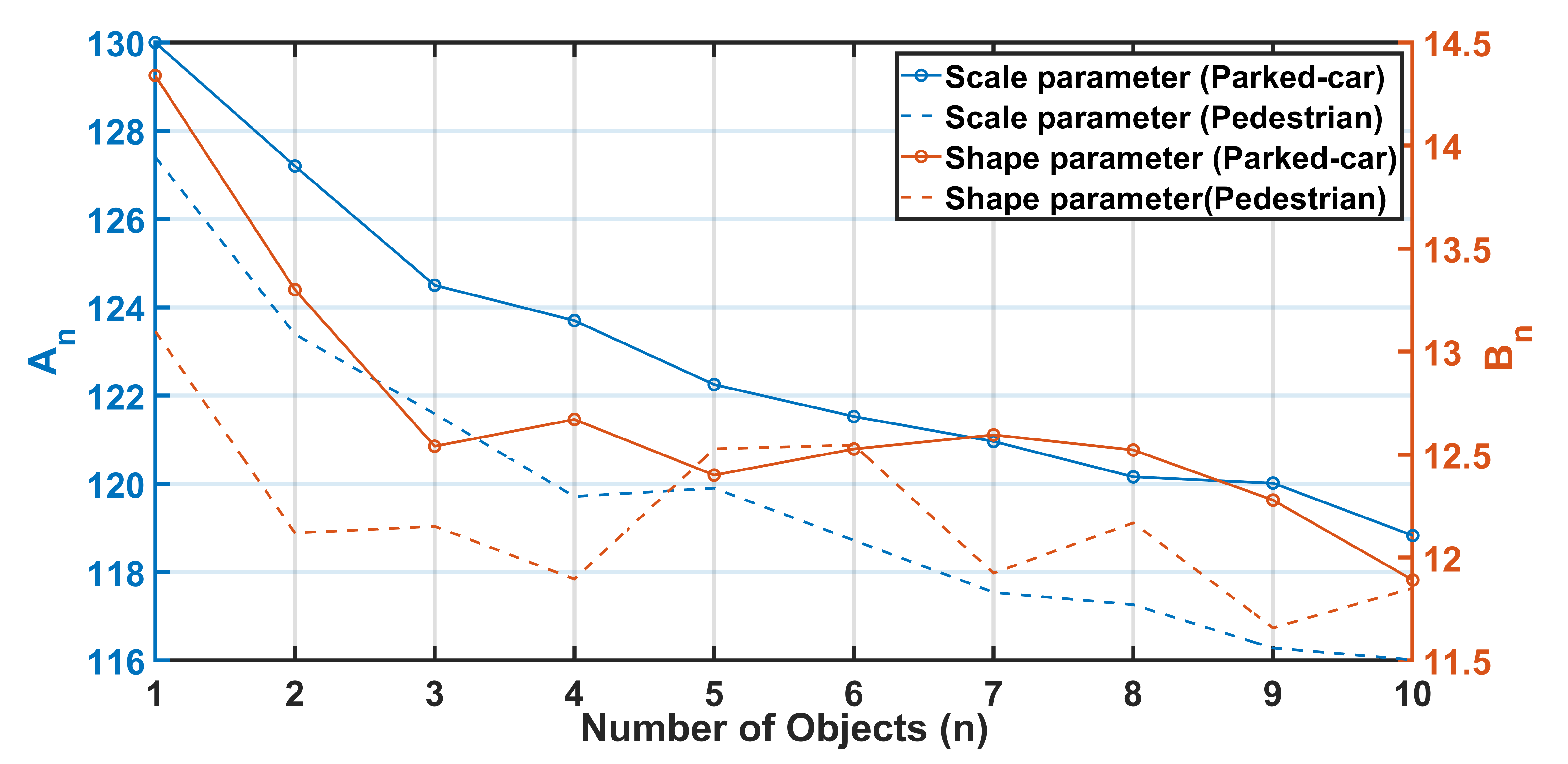}
  %\vspace{-15pt}
  \caption{Parameters of Weibull distributions of path loss for n cars/pedestrians.}\label{Curlfitingpower}
  \end{center}
\end{figure}

% \begin{table*}[]
%     \centering
%     \caption{Summary of Monte-Carlo simulation results of statistical channel parameters.}
%     \label{summaryofMonteCarlosimulation}
%     \begin{tabular}{|c|c|c|c|c|c|c|c|}

%     \end{tabular}
%     % \caption{Caption}
%     % \label{tab:my_label}
% \end{table*}

\begin{table*}[t]
\centering
\caption{Summary of Monte-Carlo simulation results of statistical channel parameters.}
\label{summaryofMonteCarlosimulation}
\begin{tabular} { |p{1cm}|p{2.5cm}|p{2.5cm}|p{2.5cm}|p{2cm}|p{2cm}|p{1.5cm}|}
% \begin{tabular}{@{} |c{1cm}|c{2.5cm}|c{2.5cm}|c{2.5cm}|c{2cm}|c{2cm}|c{1.5cm}}
 \hline
 Object & Fixed positions    & Monte-Carlo variables &Range of variables    & \makecell{ Statistic  \\  parameter} & \makecell{ Statistic \\model results  }& \makecell{ Two-sample \\Cramer-von\\ Mises \\test for a \\significance \\level of 0.01}\\
 \hline
\multirow{2}{1cm}{Pedestrian} & \multirow{2}{2.5cm}{ \makecell{ $(X_{tx},Y_{tx},Z_{tx})$  \\    $(X_{rx},Y_{rx},Z_{rx})$   }} & \multirow{2}{2.5cm}{ \makecell{ Pedestrian's  \\ location\\$(x,y,z)$ }} & \multirow{2}{2cm}{ \makecell{  $-75<x<75$\\$0<y<2$\\$14<y<16$\\$z=1$}} &  \makecell{path loss$^5$ \\due to n numbers \\of pedestrians} & \makecell{ Weibull \\distribution\\ ($A_n$ and  $B_n$) \\ see, Fig.~\ref{Curlfitingpower}} & \makecell{ Passed \\ p-value=0.46}
 \\ \cline{5-7}
 & & & & Excess time delay & \makecell{ Lognormal \\distribution\\ $\mu_n=3$\\ $\sigma_n=1$} & \makecell{ Passed \\ p-value=0.13}\\ \hline
 \multirow{2}{1cm}{Parked car} & \multirow{2}{2.5cm}{ \makecell{ $(X_{tx},Y_{tx},Z_{tx})$  \\ $(X_{rx},Y_{rx},Z_{rx})$}} & \multirow{2}{2.5cm}{\makecell{  location of \\ Parked car\\$(x,y,z)$ }} & \multirow{2}{2cm}{\makecell{  $-75<x<75$\\$y=3$\\$y=13$\\$z=1$}} & \makecell{path loss \\due to n numbers \\of parked-cars} & \makecell{ Weibull \\distribution\\ ($A_n$ and  $B_n$) \\ see, Fig.~\ref{Curlfitingpower}}  & \makecell{ Passed \\ p-value=0.4}\\ \cline{5-7}
 & & & & Excess time delay & \makecell{ Lognormal \\distribution\\ $\mu=3$\\ $\sigma=1$} & \makecell{ Passed \\ p-value=0.35}\\ \hline
%   \multicolumn{6}{l}{$^{1}$\footnotesize{  Because of the superposition assumption, each object is analyzed independently, that is, in the Monte-Carlo simulation, }  } \\
%   \multicolumn{6}{l}{ \footnotesize{  the other objects are not considered simultaneously. }  } \\
% \multicolumn{6}{l}{$^{2}$\footnotesize{See Table\ref{SimulationSetupsummary} for values.}  } \\
% \multicolumn{6}{l}{$^{3}$\footnotesize{  The distance is in meters and  path loss in [dB] and propagation time in [ns].}  } \\
% \multicolumn{6}{l}{$^{4}$\footnotesize{  The lampposts are  equidistantly   distributed over the street with fixed spacing. Each side of the street has 5 lampposts  }  } \\
% \multicolumn{6}{l}{\footnotesize{with a separation  distance of 32 m.}  } \\
\end{tabular}
\end{table*}
 \section{simulation results}
\label{SecSimResult}
The purpose of this study is to validate the quasi-deterministic ray tracing (QD-RT) method by comparing it with the deterministic ray tracing (D-RT) method for a street canyon scenario. To accomplish this, the pass loss and excess delay time distributions due to a pedestrian (parked cars) located randomly on the sidewalk (along the street) in the street canyon scenario shown in Fig.~\ref{Scenariotopology} with dimensions listed in Table~\ref{SimulationSetupsummary} are numerically calculated using both methods. The PDF of the  bistatic RCS for the pedestrians and parked cars are first obtained using the Physical Optics method, with logistic distributions observed for both cases. It is observed that the pedestrian and parked car follow the logistic distributions as listed in Table~\ref{bistaticRCSPDF}. The mean values for a car and a pedestrian are 11 and 6.17 dBsm, respectively. However, the maximum values (corresponding to the specular points) for a car and a pedestrian are around 60 dBsm and 40 dBsm, which yields a considerable difference of approximately 20 dBsm. Monte Carlo simulations are then conducted with a total of 1000 Monte-Carlo simulations with $n \in \{1, ..., 10\}$ pedestrians, randomly positioned on the sidewalks is performed with the resulting path loss distributions fitted to Weibull distributions with scale and shape parameters. Excess time delay distributions for both pedestrians and parked cars are also calculated, with lognormal distributions observed. The statistical parameters of the path loss and excess time delay distributions are presented in Fig.~\ref{Curlfitingpower} and table~\ref{summaryofMonteCarlosimulation}, respectively. This study demonstrates that the QD-RT method offers the same path loss distributions as the D-RT method with lower complexity, making it a promising approach for analyzing complex scenarios such as street canyon scenarios in mmWave wireless communication systems.

\section{Conclusion}
\label{SecConclusion}
In conclusion, the proposed quasi-deterministic ray tracing method using a statistical bistatic distribution to model the Radar Cross Section of various irregular objects showed promising results in analyzing the propagation of electromagnetic waves in street canyon scenarios. The method provided the same path loss and excess time delay distributions as the deterministic ray tracing model while offering lower complexity. The study also found that the scenario-specific PDF of bistatic RCS of irregular objects followed logistic distributions and the path loss and excess time delay followed Weibull and lognormal distributions, respectively.  This study highlights the potential of the QD-RT method for analyzing complicated scenarios, such as street canyon scenarios, in mmWave wireless communication systems.

%%%%%%%%%%%%%%%%%%%%%%%%%%%%%%%%%%%%%%%%%%%%%%%%%%%%%%%%%%%%%%%%%%%%%%%%%%%%%

\section*{acknowledgments}
The present work received funding from the European Union’s
Framework Programme for Research and Innovation Horizon
2020 under Grant Agreement No. 861222 (MINTS project).

%%%%%%%%%%%%%%%%%%%%%%%%%%%%%%%%%%%%%%%%%%%%%%%%%%%%%%%%%%%%%%%%%%%%%%%%%%%%%

\bibliographystyle{IEEEtran}

\bibliography{IEEEabrv,IEEEexample}

\end{document}